\documentstyle[12pt]{article}

\newcommand{\beqn}{\begin{equation}}
\newcommand{\eeqn}{\end{equation}}
\newcommand{\beqnarray}{\begin{eqnarray}}
\newcommand{\eeqnarray}{\end{eqnarray}}

\newcommand{\rd}{\partial}
\newcommand{\dfrac}[2]{ \frac{\displaystyle #1}{\displaystyle #2} }

\newcommand{\diag}{\mathop{{\rm diag}}}
\newcommand{\Tr}{\mathop{{\rm Tr}}}
\newcommand{\const}{{\rm const}\;}
\newcommand{\gl}{{\rm gl}}

\newcommand{\SU}{{\rm SU}}
\newcommand{\bfC}{{\bf C}}
\newcommand{\bfZ}{{\bf Z}}
\newcommand{\calF}{{\cal F}}
\newcommand{\calO}{{\cal O}}
\newcommand{\calT}{{\cal T}}
\newcommand{\mutilde}{\tilde{\mu}}
\begin{document}

\title{Dual Isomonodromic Problems and \\
Whitham Equations}
\author{Kanehisa Takasaki 
\thanks{Partly supported by the Grant-in-Aid for Scientific 
Researches, Priority Area 231 ``Infinite Analysis'', 
the Ministry of Education, Science and Culture, Japan}
\\
{\normalsize Department of Fundamental Sciences, Kyoto University}\\
{\normalsize Yoshida, Sakyo-ku, Kyoto 606, Japan}\\
{\normalsize E-mail: \tt takasaki@yukawa.kyoto-u.ac.jp}
}
\date{}
\maketitle
\bigskip

\begin{abstract}
The author's recent results on an asymptotic description of 
the Schlesinger equation are generalized to the JMMS equation. 
As in the case of the Schlesinger equation, the JMMS equation 
is reformulated to include a small parameter $\epsilon$. By 
the method of multiscale analysis, the isomonodromic problem 
is approximated by slow modulations of an isospectral problem. 
A modulation equation of this slow dynamics is proposed, and 
shown to possess a number of properties similar to the Seiberg-
Witten solutions of low energy supersymmetric gauge theories. 
\end{abstract}
\bigskip

\begin{flushleft}
KUCP-0106
\end{flushleft}
\newpage
%%%%%%%%%%%%%%%%%%%%%%%%%%%%%%%%%%%%%%%%%%%%%%%%%%%%%%%%%%%%%%%%%%%%

\section{Introduction}

This paper aims to supplement the author's recent work 
on an asymptotic description of isomonodromic problems 
\cite{bib:Takasaki-IM} in the language of Whitham equations 
\cite{bib:Whitham}. In the previous paper, the author 
considered the Schlesinger equation \cite{bib:Schlesinger}, 
and proposed a modulation equation that is expected to 
govern the Whitham dynamics of an underlying isospectral 
problem.  The modulation equation and its solutions 
exhibit remarkable similarity with the so called 
``Seiberg-Witten solutions'' of low energy supersymmetric 
gauge theories \cite{bib:SW-integrable}.  In this paper, 
we shall extend these results to the Jimbo-Miwa-M\^ori-Sato 
(JMMS) equation \cite{bib:JMMS}. The JMMS equation is the 
simplest generalization of the Schlesinger equation with 
an extra irregular singular point at infinity, and has 
found many applications in solvable lattice models, field 
theories, and random matrices 
\cite{bib:JMMS,bib:IIKS, bib:Tracy-Widom}.  
Furthermore, this equation possesses a kind of duality 
that exchanges the roles of regular singularities at 
finite points and the irregularities at infinity 
\cite{bib:Harnad-dual}.  This duality is an isomonodromic 
counterpart of a similar duality in isospectral problems 
of Garnier and Moser type \cite{bib:AHH-dual}.  A main 
concern of the following consideration is the fate of 
this duality in the modulation equation.

\section{JMMS Equation} 

In this section, we collect basic facts about 
the JMMS equation from the literature 
\cite{bib:JMMS,bib:Tracy-Widom,bib:Harnad-dual}. 

The JMMS equation governs isomonodromic deformations of 
the matrix ODE system
\beqn
    \frac{dY}{d\lambda} = M(\lambda) Y 
\eeqn
with an $r \times r$ matrix coefficient 
\beqn
    M(\lambda) = U + \sum_{i=1}^N \dfrac{A_i}{\lambda - t_i}, 
\eeqn
where $A_i$'s are $r \times r$ complex matrices, and $U_i$ 
a diagonal matrix of the form
\beqn
    U = \diag( u_1,\cdots,u_r ) 
    = \sum_{\alpha=1}^r u_\alpha E_\alpha. 
\eeqn
(Here $E_\alpha$ is the matrix whose $(\beta,\gamma)$ element 
is $\delta_{\alpha\beta}\delta_{\alpha\gamma}$.) 
$t_i$ and $u_\alpha$ are ``time variables'' of isomonodromic 
deformations.  $\lambda = t_i$ ($i = 1,\cdots,N$) are regular 
singular points, and $\lambda = \infty$ an irregular singular 
point of Poincar\'e rank 1. 
In order to avoid logarithmic terms, we assume, 
{\it throughout this paper\/}, that the eigenvalues 
$\theta_{i\alpha}$ ($\alpha = 1,\cdots,r$) of $A_i$ have 
no integer difference, and that $u_i$'s are pairwise distinct: 
\beqn
    \theta_{i\alpha} - \theta_{i\beta} \not\in \bfZ, \quad 
    u_\alpha - u_\beta \not= 0 \quad (\alpha \not= \beta). 
\eeqn
Under this assumption, isomonodromic deformations are 
generated by the PDE's 
\beqn
    \dfrac{\rd Y}{\rd t_i} = - \dfrac{A_i}{\lambda - t_i} Y, 
    \quad 
    \dfrac{\rd Y}{\rd u_\alpha} = (\lambda E_\alpha + B_\alpha) Y, 
\eeqn
where 
\beqn
    B_\alpha = - \sum_{\beta(\not=\alpha)} 
      \dfrac{E_\alpha A_\infty E_\beta + E_\beta A_\infty E_\alpha}
            {u_\alpha - u_\beta}, 
    \quad 
    A_\infty = - \sum_{i=1}^N A_i. 
\eeqn
The Frobenius integrability conditions give the so called 
``zero-curvature'' (or ``Lax'') representation of isomonodromic 
deformations.  The essential part of the zero-curvature 
representation consists of the equations 
\beqn
    \left[ \dfrac{\rd}{\rd t_i} + \dfrac{A_i}{\lambda - t_i}, 
           \dfrac{\rd}{\rd\lambda} - M(\lambda) \right] = 0, 
    \quad 
    \left[ \dfrac{\rd}{\rd u_\alpha} - \lambda E_\alpha - B_\alpha, 
           \dfrac{\rd}{\rd\lambda} - M(\lambda) \right] = 0. 
\eeqn
They give the following equations for $A_i$'s (JMMS equation): 
\beqnarray
    \dfrac{\rd A_j}{\rd t_i} &=& 
      (1 - \delta_{ij}) \dfrac{[A_i,A_j]}{t_i - t_j} 
      + \delta_{ij} \left[ U + \sum_{k(\not= i)} \dfrac{A_k}{t_i - t_k}, 
        A_j \right], 
    \nonumber \\
    \dfrac{\rd A_j}{\rd u_\alpha} &=& 
      \left[ t_j E_\alpha + B_\alpha, A_j \right]. 
\eeqnarray

The isomonodromic deformations have two sets of invariants.  
One of them consists of the eigenvalues $\theta_{j\alpha}$ 
of $A_j$'s: 
\beqn
    \dfrac{\rd \theta_{j\beta}}{\rd t_i} = 
    \dfrac{\rd \theta_{j\beta}}{\rd u_\alpha} = 0. 
\eeqn
Another set of invariants are the diagonal elements 
of $A_\infty$: 
\beqn
    \dfrac{\rd A_{\infty,\beta\beta}}{\rd t_i} = 
    \dfrac{\rd A_{\infty,\beta\beta}}{\rd u_\alpha} = 0. 
\eeqn
These invariants are nothing but characteristic 
exponents of the ODE at the singular points.

The invariance of $\theta_{i\alpha}$ is related to a 
coadjoint orbit structure behind the JMMS equation.  
Note that the right hand side is a commutator of the form 
$[\cdots,A_j]$. This implies that the motion of $A_j$ is 
constrainted to a coadjoint orbit $\calO_j$ in $\gl(r,\bfC)$; 
the eigenvalues $\theta_{j\alpha}$ are invariants that 
determine this coajoint orbit.  Thus, the JMMS equation 
is a non-autonomous dynamical system on the direct product 
$\calO_1 \times \cdots \times \calO_N$ of coajoint orbits.  
This dynamical system can be written in the Hamiltonian form 
\beqn
    \dfrac{\rd A_j}{\rd t_i} = \{A_j, H_i\}, \quad 
    \dfrac{\rd A_j}{\rd u_\alpha} = \{A_j, K_\alpha\} 
\eeqn
with the Hamiltonians 
\beqnarray
    H_i &=& \Tr \left( UA_i + 
        \sum_{j(\not= i)} \dfrac{A_i A_j}{t_i - t_j} \right), 
    \nonumber \\
    K_\alpha &=& \Tr \left( E_\alpha \sum_{i=1}^N t_i A_i + 
        \sum_{\beta(\not= \alpha)} 
           \dfrac{E_\alpha A_\infty E_\beta A_\infty}
                 {u_\alpha - u_\beta} \right). 
\eeqnarray
The Poisson bracket is the standard Kostant-Kirillov bracket 
on $\gl(r,\bfC)$: 
\beqn
    \{ A_{i,\alpha\beta}, A_{j,\rho\sigma} \} 
    = \delta_{ij} ( - \delta_{\beta\rho} A_{j,\alpha\sigma} 
                    + \delta_{\sigma\alpha} A_{j,\rho\beta} ). 
\eeqn

\section{Dual Formalism} 

In this section, we review the notion of dual isomonodromic 
problem \cite{bib:Harnad-dual}. Although we have assumed 
the genericity condition on characteristic exponents, 
the dual problem itself can be formulated in a more general 
setting.  (This is also the case for isospectral problems 
\cite{bib:AHH-dual}.)  Suppose now that the rank of $A_i$ 
takes a general value $\ell_i$.  $\ell_i$ is also a coajoint 
orbit invariant, and constant under the JMMS equation in the 
generalized sense.  

The first step towards the dual problem is to express the 
rational matrix $M(\lambda)$ as: 
\beqn
    M(\lambda) = U - {}^tG (\lambda I - T)^{-1} F, 
\eeqn
where $F = (F_{a\alpha})$ and $G = (G_{a\alpha})$ are 
$\ell \times r$ matrices, $\ell = \sum \ell_i$, and 
$T$ is a diagonal matrix of the form 
\beqn
    T = \sum_{i=1}^N t_i D_i, \quad 
    D_i = E_{\ell_1+\cdots+\ell_{i-1}+1} + \cdots + 
          E_{\ell_1+\cdots+\ell_i}. 
\eeqn
In particular, $A_i$ is written $A_i = - {}^tG D_i F$. 
$F_{a\alpha}$ and $G_{a\alpha}$ may be understood as 
canonical coordinates with the Poisson bracket 
\beqn
    \{ F_{a\alpha}, F_{b\beta} \} 
    = \{ G_{a\alpha}, G_{b\beta} \} = 0, 
    \quad 
    \{ F_{a\alpha}, G_{b\beta} \} 
    = \delta_{ab} \delta_{\alpha\beta}. 
\eeqn
The map $(F,G) \mapsto (A_1,\cdots,A_N)$ then becomes a 
Poisson map, and the JMMS equation can be derived from 
a Hamiltonian system in the $(F,G)$-space with the same 
Hamiltonians $H_i$ and $K_\alpha$. 

The dual problem is formulated in terms of the rational 
$\ell \times \ell$ matrix 
\beqn
    L(\mu) = T - G (\mu I - U)^{-1} {}^tG, 
\eeqn
which can be rewritten 
\beqn
    L(\mu) = T + \sum_\alpha \dfrac{P_\alpha}{\mu - u_\alpha} 
\eeqn
with $P_\alpha = - F E_\alpha {}^tG$.  Note that $P_\alpha$ 
is a rank 1 matrix; this restriction, too, can be relaxed 
\cite{bib:Harnad-dual,bib:AHH-dual}.  
The map $(F,G) \mapsto (P_1,P_2,\cdots)$ is a Poisson map 
for the Kostant-Kirillov bracket 
\beqn
    \{ P_{\alpha,ab}, P_{\beta,cd} \}  = \delta_{\alpha\beta} 
      ( - \delta_{bc} P_{\beta,ad} + \delta_{da} P_{\beta,cb} ). 
\eeqn
The previous Hamiltonians $H_i$ and $K_\alpha$ can also 
be viewed as a function of $P_\alpha$'s, and induces 
a non-autonomous Hamiltonian system in these new variables.  
This again gives an equation of JMMS type.  The corresponding 
isomonodromic problem can be written 
\beqn
    \frac{dZ}{d\lambda}  = - L(\mu) Z, \quad 
    \dfrac{\rd Z}{\rd u_\alpha} 
      = \dfrac{P_\alpha}{\mu - u_\alpha} Z, \quad 
    \dfrac{\rd Z}{\rd t_i} = - (\mu D_i + Q_i) Z, 
\eeqn
where 
\beqn
    Q_i = - \sum_{j(\not= i)} 
      \dfrac{D_i P_\infty D_j + D_j P_\infty D_i}{t_i - t_j}, 
    \quad 
    P_\infty  = - \sum_\alpha P_\alpha. 
\eeqn

\section{Introducing Small Parameter}

We now introduce a small parameter $\epsilon$ by the following 
substitution rule (which was first proposed by Vereschagin for 
the Painlev\'e equations \cite{bib:Vereschagin}): 
\beqnarray
    \frac{\rd}{\rd \lambda} \to \epsilon \frac{\rd}{\rd \lambda}, 
    \quad 
    \frac{\rd}{\rd t_i} \to \epsilon \frac{\rd}{\rd T_i}, 
    \quad 
    \frac{\rd}{\rd u_\alpha} \to \epsilon \frac{\rd}{\rd u_\alpha}, 
    \nonumber \\
    \frac{1}{\lambda - t_i} \to \frac{1}{\lambda - T_i}, 
    \quad 
    \sum u_\alpha E_\alpha \to \sum U_\alpha E_\alpha. 
\eeqnarray
As we shall show below, $T_i$ and $U_\alpha$ play the role of 
``slow variables'' in the terminology of multiscale analysis. 
The JMMS equation now takes the $\epsilon$-dependent form 
\beqnarray
    \epsilon \dfrac{\rd A_i}{\rd T_i} &=& 
      (1 - \delta_{ij}) \dfrac{[A_i,A_j]}{T_i - T_j} 
      + \delta_{ij}\left[ U + \sum_{k(\not= i)} 
          \dfrac{A_k}{T_i - T_k}, A_j \right], 
    \nonumber \\
    \epsilon \dfrac{\rd A_j}{\rd U_\alpha} &=& 
      \left[ T_j E_\alpha + B_\alpha, A_j \right], 
\eeqnarray
where 
\beqn
    B_\alpha = - \sum_{\beta(\not= \alpha)} 
      \dfrac{E_\alpha A_\infty E_\beta + E_\beta A_\infty E_\alpha} 
            {U_\alpha - U_\beta}. 
\eeqn

A central idea of multiscale analysis is the use of two sets 
of independent variables (``fast'' and ``slow'' variables) 
that represent two different scales of phenomena.  In the 
present setting, $A_j$'s are thus first assumed to be a 
function of ``fast variables'' $(t,u) = (t_i,u_\alpha)$ 
and ``slow variables'' $(T,U) = (T_i,U_\alpha)$, then 
restricted to the subspace 
\beqn
    t_i = \epsilon^{-1} T_i, \quad 
    u_\alpha = \epsilon^{-1} U_\alpha. 
\eeqn
Accordingly, derivatives in the original differential equation 
are replaced by the sum of derivatives in the fast and slow 
variables: 
\beqn
    \epsilon \dfrac{\rd A_j}{\rd T_i} \to 
      \dfrac{\rd A_j}{\rd t_i} 
      + \epsilon \dfrac{\rd A_j}{\rd T_i}, 
    \quad 
    \epsilon \dfrac{\rd A_j}{\rd U_\alpha} \to 
      \dfrac{\rd A_j}{\rd u_\alpha} 
      + \epsilon \dfrac{\rd A_j}{\rd U_\alpha}. 
\eeqn

Subsequent analysis is a kind of perturbation theory. 
Namely, one assumes the asymptotic form (as $\epsilon \to 0$) 
\beqn
    A_j = A_j^{(0)}(t,u,T,U) + A_j^{(1)}(t,u,T,U) \epsilon + \cdots, 
\eeqn
and considers the differential equation order-by-order in  
$\epsilon$-expansion.  The lowest order of this expansion 
for the JMMS equation yields the following equation: 
\beqnarray 
    \dfrac{\rd A_j^{(0)}}{\rd t_i} &=& 
      (1 - \delta_{ij}) \dfrac{[A_i^{(0)},A_j^{(0)}]}{T_i - T_j} 
      + \delta_{ij} \left[ U + \sum_{k(\not= i)} 
          \dfrac{A_k^{(0)}}{T_i - T_k}, A_j^{(0)} \right], 
    \nonumber \\
    \dfrac{\rd A_j^{(0)}}{\rd U_\alpha} &=& 
      \left[ T_j E_\alpha + B_\alpha^{(0)}, A^{(0)} \right]. 
\eeqnarray
Here $B_\alpha^{(0)}$ is given by the same formula 
as $B_\alpha$ except that $A_\infty$ is replaced by 
$A^{(0)}_\infty = - \sum_{i=1}^N  A_i^{(0)}$. 

Note that the lowest order equation itself does not determine 
the $(T,U)$-dependence (i.e., ``slow dynamics'') of $A_j^{(0)}$'s. 
The slow variables appear as {\it parameters} in the lowest 
order equation.  In the standard recipe of multiscale analysis, 
the slow dynamics is to be determined by the next order analysis.  
We shall, however, take a different approach to bypass technical 
difficulties (or, rather, complexity).

\section{Isospectral Problem}

Our approach to slow dynamics, which is still heuristic 
but considerably transparent, is based on the fact that 
the above lowest order equation is an isospectral problem.  
The above equation indeed has the isospectral Lax representation 
\beqnarray
    \left[ \dfrac{\rd}{\rd t_i} 
      + \dfrac{A_i^{(0)}}{\lambda - T_i}, 
        M^{(0)}(\lambda) \right] = 0, 
    \quad 
    \left[ \dfrac{\rd}{\rd u_\alpha} 
      - \lambda E_\alpha - B_\alpha^{(0)}, 
        M^{(0)}(\lambda) \right] = 0, 
\eeqnarray
where 
\beqn
    M^{(0)}(\lambda) = 
      U + \sum_{i=1}^N \dfrac{A_i^{(0)}}{\lambda - T_i}. 
\eeqn
In particular, the characteristic polynomial 
$\det\Bigl( \mu I - M^{(0)}(\lambda) \Bigr)$ 
is constant under the $(t,u)$-flows.  It however depends on 
$(T,U)$. Thus the slow dynamics in $(T,U)$ may be described 
as slow deformations of the characteristic polynomial or 
of the spectral curve 
\beqn
    \det\Bigl( \mu I - M^{(0)}(\lambda) \Bigr) = 0. 
\eeqn

Spectral curves of the above type has been studied in detail 
in the context of isospectral problems of Garnier and Moser 
\cite{bib:AHHP,bib:Beauville}.  A fundamental result is that 
the spectral curve $C_0$ on the $(\lambda,\mu)$-plane can be 
compactified to a nonsingular curve $C$.  This is due to the 
previous assumption on the eigenvalues of $A_i$'s and $U$; 
if they have multiple eigenvalues, $C$ has nodal singularities. 
Let $\pi$ denote the projection $\pi(\lambda,\mu) = \lambda$ 
from $C_0$ to the punctured Riemann sphere 
$\bfC P^1 \setminus \{T_1,\cdots,T_N,\infty\}$. This extends 
to $C$ and gives an $r$-fold ramified covering of $C$ over 
$\bfC P^1$. $C$ is obtained from $C_0$ by adding several 
points to the holes over $\lambda = T_1,\cdots,T_N,\infty$.  
The holes over $\lambda = \infty$ are filled by the $r$ points 
$(\lambda,\mu) = (\infty,U_\alpha)$ ($\alpha = 1,\cdots,r$).  
The situation of holes over $\lambda = T_i$ becomes clearer 
in terms of the new coordinate 
\beqn
    \mutilde = f(\lambda) \mu, \quad 
    f(\lambda) = \prod_{i=1}^N (\lambda - T_i), 
\eeqn
in place of $\mu$. The equation of the spectral curve can be 
rewritten 
\beqn
    F(\lambda,\mutilde) 
    = \det\Bigl(\mutilde I - f(\lambda) M^{(0)}(\lambda) \Bigr) 
    = 0. 
\eeqn
Compactification is achieved by adding the $r$ points 
$(\lambda,\mutilde)  = (T_i, f'(T_i)\theta_{i\alpha})$ 
($\alpha = 1,\cdots,r$). The compactified spectral curve 
has the genus 
\beqn
    g = \frac{1}{2} (r - 1)(rN - 2). 
\eeqn

In order to describe slow dynamics, one has to choose 
a suitable system of parameters (``moduli'')  in the 
space (``moduli space'') of all possible spectral curves. 
Slow dynamics can be formulated in the form of differential 
equations (``modulation equations'') for those moduli 
with respect to the slow variables $(T,U)$. Note that 
the isomonodromic invariants $\theta_{i\alpha}$ and 
$A_{\infty,\alpha\alpha}$ are also constants of slow dynamics.  

Such a set of moduli can be singled out from the modified 
characteristic polynomial $F(\lambda,\mutilde)$. 
$F(\lambda,\mutilde)$ turns out to have an expansion 
of the form 
\beqn
    F(\lambda,\mutilde) = F^0(\lambda,\mutilde) 
      + f(\lambda) \sum_{s=2}^r \sum_{m=0}^{\delta_s - 1} 
         h_{ms} \lambda^m \mutilde^s, 
\eeqn
where $\delta_s$ is given by 
\beqn
    \delta_s = (N-1)s - N. 
\eeqn
$F^0(\lambda,\mutilde)$ is a polynomial whose 
coefficients are determined by the isomonodromic invariants 
and $(T,U)$ only.  The coefficients $h_{\delta_s-1,s}$ 
($2 \le s \le r$), too,  are comprised of these quantities 
only.  The remaining coefficients $h_{ms}$ ($2 \le s \le r$, 
$0 \le m \le \delta_s - 2$) give the moduli in the above sense.  
Thus, as in the case of moduli spaces of Seiberg-Witten type 
\cite{bib:SW-integrable}, we have altogether $g$ moduli: 
\beqn 
    \sum_{s=2}^r (\delta_s - 1) = g. 
\eeqn

Thus, the leading order term $A^{(0)}_i$ of the multiscale expansion 
turns out to be a solution of the isospectral problem with slowly 
varying spectral invariants $h_{ms} = h_{ms}(T,U)$.  Since this 
kind of isospectral problems can be solved by Abelian (or Riemann 
theta) functions \cite{bib:AHHP,bib:Beauville}, the multiscale 
expansion $A_i = A^{(0)}_i + O(\epsilon)$ gives an asymptotic 
expression of JMMS transcendents as a modulated Abelian 
functions.

\section{Modulation Equation}

The first step for deriving the modulation equation is 
the following WKB ansatz (which dates back to Flaschka and 
Newell's early work on isomonodromic problems 
\cite{bib:Flaschka-Newell}): 
\beqn
    Y = \Bigl(\phi^{(0)}(t,u,T,U,\lambda) 
         + \phi^{(1)}(t,u,T,U,\lambda) \epsilon + \cdots \Bigr) 
        \exp \epsilon^{-1} S(T,U,\lambda). 
\eeqn
Here $Y$ and $\phi^{(k)}$'s are understood to be vector valued; 
$S$ is a scalar phase function. Note that whereas $\phi^{(k)}$'s 
depend on both the fast and slow variables, $S$ is assumed to 
be a function of slow variables only.  Inserting the above ansatz 
into the isomonodromic linear problem and expanding in powers of 
$\epsilon$ give a series of differential equations for 
$\phi^{(k)}$'s and $S$.

The lowest order equation reads: 
\beqnarray
    \frac{\rd S}{\rd \lambda} \phi^{(0)} 
      &=& M^{(0)}(\lambda) \phi^{(0)}, 
    \nonumber \\
    \dfrac{\rd \phi^{(0)}}{\rd t_i} + \dfrac{\rd S}{\rd T_i} \phi^{(0)} 
      &=& - \dfrac{\rd A^{(0)}}{\lambda - T_i} \phi^{(0)}, 
    \nonumber \\
    \dfrac{\rd \phi^{(0)}}{\rd u_\alpha} 
      + \dfrac{\rd S}{\rd U_\alpha} \phi^{(0)}
      &=& (\lambda E_\alpha + B_\alpha^{(0)}) \phi^{(0)}. 
\eeqnarray 
(Note that these equations determine $\phi^{(0)}$ up to 
multiplication $\phi^{(0)} \mapsto \phi^{(0)} h$ by a 
scalar function $h = h(T,U)$ of slow variables.) 
These equations can be rewritten into the isospectral 
linear problem 
\beqn
    \mu \psi = M^{(0)}(\lambda) \psi, 
    \quad 
    \dfrac{\rd \psi}{\rd t_i}  
      = - \dfrac{A_i^{(0)}}{\lambda - T_i} \psi, 
    \quad 
    \dfrac{\rd \psi}{\rd u_\alpha} 
      = (\lambda E_\alpha + B_\alpha^{(0)}) \psi 
\eeqn
by change of variables as 
\beqn
    \mu = \frac{\rd S}{\rd \lambda}, 
    \quad 
    \psi = \phi^{(0)} \exp \Bigl( 
           \sum t_i \dfrac{\rd S}{\rd T_i} + 
           \sum u_\alpha \dfrac{\rd S}{\rd U_\alpha} \Bigr). 
\eeqn
In particular, the characteristic equation 
$\det \Bigl(\mu I - M^{(0)}(\lambda) \Bigr) = 0$ is 
satisfied by $\mu = \rd S/\rd \lambda$. 

We derive the modulation equation by identifying this $\psi$ 
with the algebro-geometric expression of $\psi$ in terms of 
Baker-Akhiezer functions (with a suitable correction by a 
scalar factor $h$ such that $\phi = \phi^{(0)} h$).  In the 
present case, such an expression takes the form \cite{bib:AHHP} 
\beqn
    \psi = \phi \exp \Bigl( \sum t_i \Omega_{T_i} + 
           \sum u_\alpha \Omega_{U_\alpha} \Bigr), 
\eeqn
where $\phi$ is a vector valued function made from various 
Riemann theta functions on the spectral curve, and $\Omega_{T_i}$ 
and $\Omega_{U_\alpha}$ are primitive functions of meromorphic 
differentials $d\Omega_{T_i}$ and $\Omega_{U_\alpha}$, 
\beqn
    \Omega_{T_i} = \int^{(\lambda,\mu)} d\Omega_{T_i}, \quad 
    \Omega_{U_\alpha} = \int^{(\lambda,\mu)} d\Omega_{U_\alpha}. 
\eeqn
[More precisely, such an expression of Baker-Akhiezer functions 
is available upon fixing a symplectic basis $A_I,B_I$ 
($I = 1,\cdots,g$) on the spectral curve.  In the present 
situation, the symplectic basis has to be chosen so that 
the above WKB solution gives a correct approximation to 
the isospectral problem; this is a nontrivial problem. 
In the following, we {\it assume} that such a suitable 
symplectic basis is selected.]  The meromorphic differentials 
$d\Omega_{T_i}$ and $d\Omega_{U_\alpha}$ are characterized 
by the following conditions: 
\begin{itemize}
\item 
  $d\Omega_{T_i}$ has poles at the $r$ points in $\pi^{-1}(T_i)$ 
  and is homomorphic outside $\pi^{-1}(T_i)$ . Its singular 
  behavior at these points is such that 
\beqn
    d\Omega_{T_i} = - d\mu + \mbox{non-singular}.  
\eeqn
\item 
  $d\Omega_{U_\alpha}$ has poles at the $r$ points in 
  $\pi^{-1}(\infty)$ and is holomorphic outside 
  $\pi^{-1}(\infty)$ . Its singular behavior at these points 
  is such that 
\beqn
    d\Omega_{U_\alpha} = d\lambda + \mbox{non-singular}. 
\eeqn
\item
  $d\Omega_{T_i}$ and $d\Omega_{U_\alpha}$ have vanishing 
  $A_I$-periods: 
\beqn
    \oint_{A_I} d\Omega_{T_i} = 
    \oint_{A_I} d\Omega_{U_\alpha} = 0 \quad 
    (I = 1,\cdots,g). 
\eeqn
\end{itemize}
Matching the exponential part in the two expressions of $\psi$, 
we obtain the equations 
\beqn
    \dfrac{\rd S}{\rd T_i} = \Omega_{T_i}, \quad 
    \dfrac{\rd S}{\rd U_\alpha} = \Omega_{U_\alpha}. 
\eeqn
This is an expression of our modulation equation.

This modulation equation can be rewritten 
\beqn
    \left. \frac{\rd}{\rd T_i} dS \right|_{\lambda=\const} 
       = d\Omega_{T_i}, 
    \quad 
    \left. \frac{\rd}{\rd U_\alpha} dS \right|_{\lambda=\const}
       = d\Omega_{U_\alpha} 
\eeqn
in terms of the meromorphic differential 
\beqn
    dS = \mu d\lambda. 
\eeqn
Here, as in the literature of Whitham equations 
\cite{bib:Krichever-Whitham,bib:Dubrovin-Whitham}, 
``$\lambda=\const$'' stands for differentiation while leaving 
$\lambda$ constant. In particular, the following equation of 
Flaschka-Forest-McLaughlin type \cite{bib:FFM} are satisfied 
under the above equations: 
\beqnarray 
    \left. \dfrac{\rd}{\rd T_i} d\Omega_{T_j} 
    \right|_{\lambda=\const} 
    = \left. \dfrac{\rd}{\rd T_j} d\Omega_{T_i} 
      \right|_{\lambda=\const}, 
    &\quad& 
    \left. \dfrac{\rd}{\rd U_\alpha} d\Omega_{U_\beta} 
    \right|_{\lambda=\const} 
    = \left. \dfrac{\rd}{\rd U_\beta} d\Omega_{U_\alpha} 
      \right|_{\lambda=\const}, 
    \nonumber \\ 
    \left. \dfrac{\rd}{\rd T_i} d\Omega_{U_\alpha} 
    \right|_{\lambda=\const} 
    &=& \left. \dfrac{\rd}{\rd U_\alpha} d\Omega_{T_i} 
      \right|_{\lambda=\const}. 
\eeqnarray
Thus our modulation equation turns out to give a special 
family of Whitham equations.  These equations, though rather 
indirectly, determine the slow dynamics of the spectral curve 
(hence of the spectral invariants $h_{ms} = h_{ms}(T,U)$) 
for the isospectral aproximation $A_i \sim A^{(0)}_i$ to the 
JMMS equation. 

The same multiscale argument applies to the dual isomonodromic 
problem, too.  The WKB ansatz now takes the form 
\beqn 
    Z = \Bigl( \chi^{(0)}(t,u,T,U,\mu) 
          + \chi^{(1)}(t,u,T,U,\mu) \epsilon 
          + \cdots \Bigr) \exp \epsilon^{-1} \Sigma(T,U,\mu).
\eeqn
This leads to slow dynamics of the dual isospectral problem 
\cite{bib:AHH-dual} based on the dual expression 
\beqn
    \det \Bigl( \lambda I - L^{(0)}(\mu) \Bigr) = 0 
\eeqn
of the same spectral curve.  The modulation equation is obtained  
in the dual form
\beqn
    \left. \dfrac{\rd}{\rd T_i} d\Sigma  \right|_{\mu=\const} 
      = d\Omega_{T_i}, 
    \quad 
    \left. \dfrac{\rd}{\rd U_\alpha} d\Sigma \right|_{\mu=\const}
      = d\Omega_{U_\alpha}, 
\eeqn
where 
\beqn
    d\Sigma = - \lambda d\mu. 
\eeqn
It should be noted that a similar dual representation is known 
for the Seiberg-Witten solutions \cite{bib:SW-integrable}.

\section{Period Map and Prepotential}

The above modulation equation, like the modulation equation 
of the Schlesinger equation \cite{bib:Takasaki-IM}, can be 
treated as a variant of problems of Seiberg-Witten type.  
We show a list of main results below.  

1. {\it General solutions of the modulation equation are 
obtained by an inverse period map\/}.  Let us consider 
the $g$-dimensional period map $h = (h_{ms}) \mapsto a = (a_I)$ 
given by the $A_I$-periods
\beqn
    a_I = \oint_{A_I} dS \quad (I = 1,\cdots,g). 
\eeqn
On very general grounds \cite{bib:Takasaki-IM}, 
one can prove that the Jacobian of this map is non-vanishing. 
The slow variables $(T,U)$ are included in the period integrals 
as parameters. Accordingly, the the inverse period map should 
be written $h = h(T,U,a)$.  With $a$ being fixed, this gives 
a family of deformations of the spectral curve, which turns out 
to solve the above modulation equation.  Varying $a$, one 
obtains a general solution.  

2. {\it With $(T,U)$ being fixed, this inverse period map 
yields deformations of Seiberg-Witten type\/}.  This means 
that the meromorphic differential $dS$ obeys the deformation 
equation
\beqn
    \left. \dfrac{\rd}{\rd a_I} dS \right|_{\lambda=\const}
    = d\omega_I  \quad 
    (I = 1,\cdots,g),  
\eeqn
where $d\omega_I$'s are a basis of holomorphic differentials 
on the spectral curve normalized as
\beqn
    \oint_{A_J} d\omega_I = \delta_{IJ}. 
\eeqn

3. {\it A prepotential of Seiberg-Witten type exists\/}.  
The prepotential $\calF = \calF(T,U,a)$ can be defined as 
a (non-zero) solution of the following equations: 
\beqn
    \dfrac{\rd \calF}{\rd a_I} = b_I, \quad 
    \dfrac{\rd \calF}{\rd T_i} = H_i, \quad 
    \dfrac{\rd \calF}{\rd U_\alpha} = K_\alpha. 
\eeqn
Here $b_I$ denotes the $B_I$-period of $dS$, 
\beqn
    b_I = \oint_{B_I} dS, 
\eeqn
and $H_i$ and $K_\alpha$ are the Hamiltonians already 
introduced.  In particular, the second derivatives of $\calF$ 
coincide with the matrix elements of the period matrix: 
\beqn
    \frac{\rd^2 \calF}{\rd a_I \rd a_J} = \calT_{IJ} 
    = \oint_{B_J} d\omega_I. 
\eeqn

\section{Conclusion} 

We have shown that the previous work on the Schlesinger equation 
can be generalized to the JMMS equation.  Although the derivation 
of the modulation equation is still heuristic, the modulation 
equation has turned out to possess a number of remarkable 
properties that strongly suggest that our modulation equation 
is a correct one.  We have also seen that the duality structure 
of the JMMS equation is inherited to the modulation equation. 

A straightforward generalization of the present situation 
is to add one more irregular singular point at, say, 
$\lambda = 0$.  The rational matrix $M(\lambda)$ then 
takes the form
\beqn 
    M(\lambda) = U + \sum \dfrac{A_i}{\lambda - T_i} 
      + V \lambda^{-2}. 
\eeqn
In this kind of isomonodromic problems, $U$ and $V$ are 
usually required to be semi-simple. The beautiful duality 
in the JMMS equation, however, then ceases to exist.  
Relaxing this requirement is an interesting issue.  
A similar situation takes place in the $N$-periodic 
Toda chain; it has an $N \times N$ Lax matrix of the form 
$L = U + W \lambda^{-1} + V \lambda^{-2}$ with $U$ and 
$V$ being nilpotent.  It is well known that a dual 
representation of this case is provided by a $2 \times 2$ 
monodromy matrix $M$ on the lattice. This fact seems to 
suggest that a similar duality can exist in isomonodromic 
problems with two irregular singular points, too. 

Another intriguing issue is to extend the present 
consideration to the case of an elliptic matrix.  
Remarkably, our $M$-matrix has almost the same structure 
as the Lax operator of the XXX Gaudin model 
\cite{bib:Sklyanin-Gaudin}.  Since an elliptic analogue 
of the XXX Gaudin model is given by the XYZ Gaudin model 
\cite{bib:Sklyanin-Takebe}, It is natural to expect 
that there will be an elliptic version of the Schlesinger 
or JMMS equation.  In the case of $2 \times 2$ matrix, 
such an elliptic isomonodromic problem will be equivalent 
to Okamoto's isomonodromic problem on a torus 
\cite{bib:Okamoto-torus}. The corresponding isospectral 
problem and its Whitham modulations should be described 
by a spectral cover of the torus. This situation, too, 
is interesting in the context of the Seiberg-Witten 
solutions, because the Seiberg-Witten solution of 
$N = 4$ supersymmetric $\SU(2)$ gauge theory is also 
related to a spectral cover of the torus 
\cite{bib:SW-integrable}.


\begin{thebibliography}{99}

\bibitem{bib:Takasaki-IM}
K. Takasaki, 
Spectral curves and Whitham equations in 
isomonodromic problems of Schlesinger type, 
solv-int/9704004.

\bibitem{bib:Whitham}
G.B. Whitham, 
Linear and Nonlinear Waves 
(Wiley-Interscience, New York, 1974). 

\bibitem{bib:Schlesinger}
L. Schlesinger, 
\"Uber eine Klasse von Differentialsystemen beliebiger 
Ordnung mit festen kritischen Punkten, 
J. f\"ur Math. {\bf 141} (1912), 96-145. 

\bibitem{bib:SW-integrable}
H. Itoyama and A. Morozov, 
Integrability and Seiberg-Witten theory: Curves and periods,  
Nucl. Phys. {\bf B477} (1996), 855-877; 
%% hep-th/9511126 
Prepotential and the Seiberg-Witten theory, 
hep-th/9512161. 
\newline
A. Marshakov, 
On integrable systems and supersymmetric gauge theories, 
hep-th/9702083. 

\bibitem{bib:JMMS}
M. Jimbo, T. Miwa, Y. M\^ori and M. Sato, 
Density matrix of an impenetrable Bose gas and
the fifth Painlev\'e transcendents, 
Physica {\bf 1D} (1980), 80-158. 

\bibitem{bib:IIKS}
A.R. Its, A.G. Izergin, V.E. Korepin and N.A. Slavnov, 
Differential equations for quantum correlation functions, 
Int. J. Mod. Phys. {\bf B4} (1990), 1003-1037. 

\bibitem{bib:Tracy-Widom}
C. Tracy and H. Widom, 
Introduction to random matrices, 
Lecture Notes in Phys. {\bf 424}, pp. 103-130 
(Springer-Verlag, Berlin, Heidelbeg, New York, 1993). 
%% hep-th/9210073. 

\bibitem{bib:Harnad-dual}
J. Harnad, 
Dual isomonodromic deformations and moment maps 
to loop algebras, 
Commun. Math. Phys. {\bf 166} (1994), 337-365. 

\bibitem{bib:AHH-dual}
M.R. Adams, J. Harnad and J. Hurtubise, 
Dual moment maps into loop algebras, 
Lett. Math. Phys. {\bf 20} (1990), 299-308. 

\bibitem{bib:Vereschagin}
V.L. Vereschagin, 
Asymptotics of solutions of the discrete string equation, 
Physica {\bf D95} (1996), 268-282; 
Nonlinear quasiclassics and Painlev\'e equations, 
hep-th/9605092. 

\bibitem{bib:AHHP}
M.R. Adams, J. Harnad and E. Previato, 
Isospectral Hamiltonian flows in finite and 
infinite dimensions, I, 
Commun. Math. Phys. {\bf 117} (1988), 451-500. 
\newline
M.R. Adams, J. Harnad and J. Hurtubise, 
{\it ditto} II, 
Commun. Math. Phys. {\bf 134} (1990), 555-585. 

\bibitem{bib:Beauville}
A. Beauville, 
Jacobiennes des courbes spectrales et syst\`emes 
hamiltoniens compl\`etement int\'egrables, 
Acta Math. {\bf 164} (1990), 211-235. 

\bibitem{bib:Flaschka-Newell}
H. Flaschka and A. Newell, 
Multiphase similarity solutions of integrable 
evolution equations, 
Physica {\bf 3D} (1981), 203-221. 

\bibitem{bib:Krichever-Whitham}
I.M. Krichever, 
The dispersionless Lax equations and 
topological minimal models, 
Commun. Math. Phys. {\bf 143} (1992), 415-429; 
The $\tau$-function of the universal Whitham hierarchy, 
matrix models and topological field theories, 
Comm. Pure. Appl. Math. {\bf 47} (1994), 437-475. 

\bibitem{bib:Dubrovin-Whitham}
B.A. Dubrovin, 
Hamiltonian formalism of Whitham hierarchies and 
topological Landau-Ginzburg models, 
Commun. Math. Phys. {\bf 145} (1992), 195-207;
Geometry of 2D topological field theories, 
Lecture Notes in Math. {\bf 1620}, pp. 120-348 
(Springer-Verlag, Berlin, Heidelbeg, New York, 1996). 
%%% hep-th/9407018. 

\bibitem{bib:FFM} 
H. Flaschka, M.G. Forest and D.W. McLaughlin,
Multiphase averaging and the inverse spectral solution 
of the Korteweg-de Vries equation, 
Comm. Pure  Appl. Math. {\bf 33} (1980) 739-784. 

\bibitem{bib:Sklyanin-Gaudin}
E.K. Sklyanin, 
Separation of variables in the Gaudin model, 
J. Soviet Math. {\bf 47} (1989), 2473-2488. 

\bibitem{bib:Sklyanin-Takebe}
E.K. Sklyanin and T. Takebe, 
Algebraic Bethe Ansatz for XYZ Gaudin model, 
q-alg/9601028. 

\bibitem{bib:Okamoto-torus}
K. Okamoto, 
On Fuchs's problem on a torus, I, 
Funkcial. Ekvac. {\bf 14} (1971), 137-152; 
{\it ditto} II, 
J. Fac. Sci. Univ. Tokyo, Sect. IA, {\bf 24} (1977), 357-371; 
D\'eformation d'une \'equation differ\'entielle lin\'eaire
avec une singularit\'e irr\'eguli\`ere sur un tore, 
{\it ibid\/}, {\bf 26} (1979), 501-518. 

\end{thebibliography}
\end{document}